# PROSPECTS OF NUCLEAR PROPULSION FOR SPACE PHYSICS

Giovanni M. Piacentino (INFN-Pisa, Fermilab Visitor)

## Scientific Case

The Solar and Space Physics (SSP) decadal survey was released in 2003 by the National Research Council. It was aimed to define and address, in a decade, several scientific questions ranging from understanding the space environment of the Earth and other solar system bodies and their reaction to external perturbation, to the study of the interstellar medium in the range from 10 to 2000 astronomical units (AU).

At present multiple observations by the Voyager spacecraft of the outer heliosphere are consistent with solar wind termination shocks, but the lack of magnetic and plasma wave signature, although composition and particle spectra are consistent with a shocked solar wind crossing, has led to controversy about the interpretation of such observations.

The still mysterious aspect of the external heliosphere dynamic emphasizes the need of studying and understanding how the heliosphere interacts with the interstellar medium in the region within some 2,000 AU from the Sun. At shorter distance it is still unknown how ejections of energy and matter from the Sun propagate, causing large fluctuations in the radiation environments of the various planets including the Earth. Without the control of space radiation hazards, extended voyages by crews beyond low Earth orbit may be very difficult and dangerous. The Interstellar Probe, in several versions, has been for a long time the NASA leading project in this field.

This mission would cross the solar system exploring the charged particle flux, the magnetic field intensity, the residual gas density and the dust in the external galactic space. The following table presents the history of the project whose principal scientific goals are to explore:

• *the outer heliosphere and the nature of its   boundaries;*
• *the outer solar system in search of clues to its origin;*
• *the interaction of the solar system with the interstellar medium;*
• *the nature of the nearby interstellar medium.*

| Mission | Propulsion System |
| --- | --- |
| Interstellar Precursor | Nuclear-electric system to 400+ AU |
| Thousand Astronomical Units | Nuclear-electric system to 1,000 AU |
| Interstellar Probe (NASA's 1990 Space Physics Roadmap) | Chemical system sending a 1,000-kg spacecraft to 200 AU using powered solar flyby |
| Interstellar Probe (NASA's 1994 Space Physics Roadmap) | Chemical system sending a small spacecraft to 200 AU |
| Interstellar Probe (NASA's 1999 Space Physics Roadmap) | Solar-sail system to 200 AU |
| Realistic Interstellar Explorer | Jupiter flyby and use of a solar-thermal propulsion system at 4 solar radii to send a small payload to 1,000 AU in <50 years |
| Innovative Interstellar Explorer | Ion propulsion powered by radioisotope power systems used to send a small payload to 200 AU in 30 years |

In the last two projects the probes need respectively a mean speed of 340.000 and 115.000 Km/h. At present no technology but a nuclear one can provide accelerations of the order of 0.5-03 g over a period of 30 – 50 years. At this stage a feasible and productive project could be that of an Interstellar Observatory. In this view, a mother vehicle with nuclear propulsion could release several small instrumental vehicles along the way in order to monitor the environment at several distances and in different directions from the Sun.

The Interstellar Observatory would also be of practical utility to manned missions in the inner/mid part of the solar system. We know how galactic cosmic radiation is a ubiquitous hazard for all voyagers beyond near-Earth space. Outside the protective shield of the Earth's magnetic field and atmosphere, astronauts are exposed to intense cosmic ray fluencies. Crew members would take near their lifetime maximum allowable dose of radiation on a multiyear round-trip mission to Mars.

The flux of galactic cosmic rays waxes and wanes following the Sun's magnetic cycle, but we still do not know how the conditions in the local interstellar medium influence the dynamical structure of the heliosphere. The Interstellar Observatory could help developing appropriate strategies in order to mitigate galactic cosmic radiation hazards. Galactic cosmic radiation also has implications for global sustainability. Changes in the conditions of the local interstellar medium might affect the Earth's climate and human radiation exposure on its surface. The known increases in cosmic ray intensity episodes are strongly associated with changes in the solar activity (e.g., the Maunder minimum) which in turn affect the terrestrial climate.

However it is unknown whether or not these variations were caused by changes in the local interstellar medium. The Interstellar Observatory will drive understanding of the effects of the local interstellar medium on heliosphere structure and its shielding of galactic cosmic rays.

Nuclear Solutions

As can be easily argued by the previous considerations, both scientific missions in deep space and Human exploration of the inner or intermediate solar system need a new kind of propulsion with very high specific impulse and the capability of working for very long periods. Till now tree different possibilities have been studied:

• *The Orion or Orion-like projects;*
• *The Ion Thruster option;*
• *The thermal nuclear reactor engine option.*

The Orion Project:

The Orion project, first proposed by Stanislaw Ulam as early as 1947, was never implemented. It would work dropping little thermonuclear or fission explosives out of the rear of the vehicle and detonating them at ~60 meters from his rear shield, an heavy and thick steel platter designed in order to catch the blast and push forward the vehicle.
Large multi-story high shock absorbers (pneumatic springs) would absorb the impulse from the plasma wave as it hit the pusher plate, spreading the millisecond shock wave over several seconds and thus giving an acceptable ride.

Two sets of shock absorption systems were tuned to different frequencies to avoid resonance. Polyethylene masses, garbage and sewage were all considered for use as reaction mass.

The smallest 4000 ton model planned for ground launch from Jackass Flats, Nevada, had each blast add 50 km/h to the craft's velocity. A graphite based oil was to be sprayed on the pusher plate before each explosion to prevent ablation of the pusher plate. The sequence would be repeated thousands of times.

Orion's potential performance was stunning, at least compared to today's chemical or even other nuclear designs. Freeman Dyson quoted that a single mission could have established a large permanent Moon base and that an Orion vehicle could reach Pluto and return to Earth in less than a year. If the potential performances are very attracting, on the contrary the difficulties of the realization are seriously discouraging. To rule and manage, not one but a long series of nuclear explosions is probably impossible even for technologies of the very distant future.

The Ion Thruster Option:

The principle of ion thrusters is due to Hermann Oberth and was published in the 1929 work "Wege zur Raumschiffahrt". In an ion thruster the ions of a charged gas are accelerated electrically in the direction opposite to the one of the vehicle by means of an electrostatic or electrodynamics device thus, in principle, using all the energy of the propellant to push forward the spacecraft. The specific impulse that can be reached is indeed very high because the speed of the ions can be relativistic. The quantity of propellant needed is so very limited and light.

The first working ion thruster was built by Harold R. Kaufman in 1959 at NASA. Just a griddled electrostatic ion thruster, this model used mercury as its fuel. During the '60 the engine was tested in suborbital flights aboard the Space Electric Rocket Test 1 and successfully operated for 31 minutes.

In spite of the mentioned very high specific thrust, the total thrust of this kind of engine is very limited because the mass of current electric power units is strongly correlated with the amount of power produced. These ion thrusters are unsuitable for launching spacecraft into orbit from the Earth, but are ideal for in-space propulsion of small vehicles where power is produced by the decay of radioactive isotopes.

The NTR Option:

NTR/NRP (Nuclear Thermal Rocket/Propulsion) engines are the best candidates to support deep interstellar medium studies and vehicles as well as drive human missions to Mars and beyond.
The NTR is the leading propulsion option for future human and instrumental exploration missions because of its high specific impulse capability (Isp ~875 to 950 sec) and attractive engine thrust-to-weight ratio (>3). The engine consists of dedicated nuclear reactor heating hydrogen to very high temperatures ~2800 K. The hot hydrogen is then emitted by a nozzle producing the desired thrust. Because only a relatively small amount of enriched uranium-235 fuel is consumed in an NTR during the primary propulsion maneuvers of a typical Mars mission, engines configured for both propulsive thrust and modest power generation (referred to as bimodal operation) provide the basis for a robust, power-rich stage with extremely efficient propulsive capability. A spacecraft can be schematized as in figure 1.

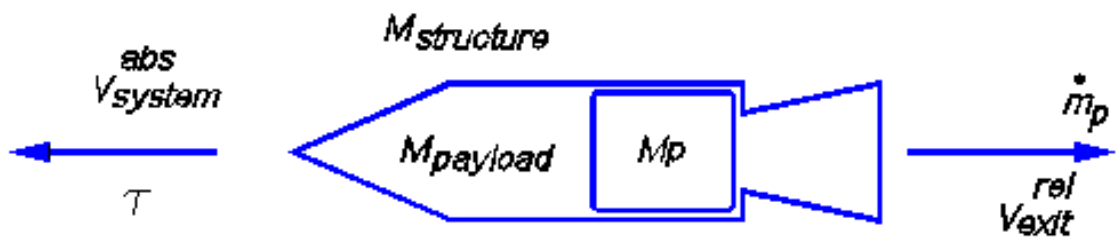

Fig 1.

where T is the thrust, $M_P$ is the mass of the ejected material and the other symbols are obvious. The spacecraft is governed by Newton equation:

$$\frac{dM\vec{V}^{abs}_{sys}}{dt} = -\dot{m}_p \vec{V}^{abs}_{exit} - A_{exit} P_{exit} \simeq -\dot{m}_p \vec{V}^{abs}_{exit} = \frac{dM}{dt} \vec{V}^{abs}_{exit}$$

the thrust and the specific impulse being respectively:

$$T \approx \dot{m}_p \vec{V}^{rel}_{exit} \qquad\qquad I_{sp} \simeq \vec{V}^{rel}_{exit}$$

Obviously the needed mass of the ejected material is:

$$M_P = (M_{stru} + M_{Payload})\left(e^{\frac{\Delta V}{I_{sp}}} - 1\right)$$

For the conservation of the mass (the mass is not addictive relativistically but not in this case) we have:

$\dot{m} = \rho \Delta V +$ SONIC conditions at throat

The state equation is thus:

$$A^* \propto \frac{\dot{m}}{P} \sqrt{T}$$

The nozzle dimensions and performance depend on the thermo-fluid-dynamics.
The thermodynamic of a NTR is ruled by the following

energetic balance:

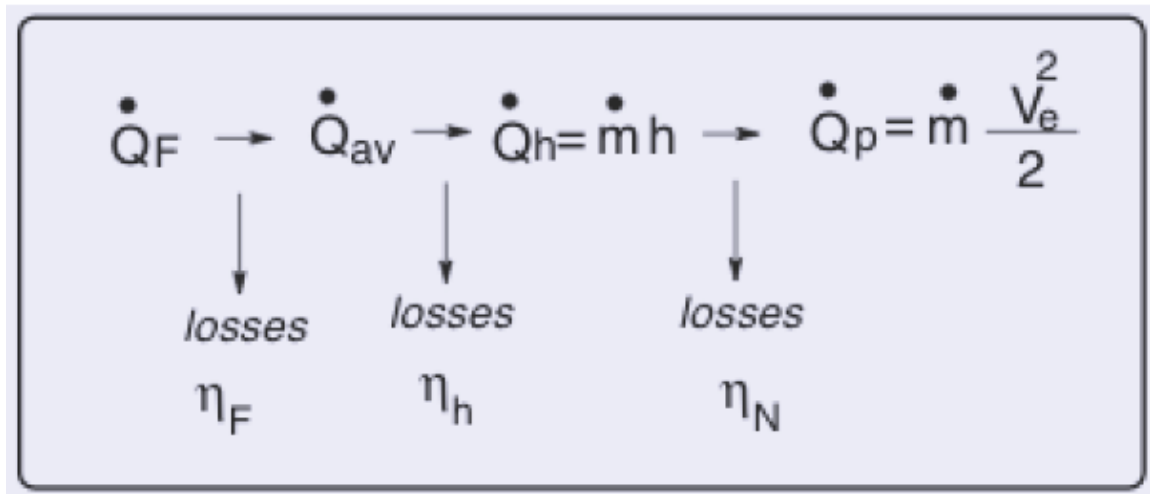

$\dot{Q}_F$ = power from fuel
$\dot{Q}_{av} = (\eta_F \dot{Q}_F)$ = energy in the propellent
$\dot{m}_p h = (\eta_F \eta_h \dot{Q}_F)$ = fraction converted into propellent enthalpy
$\frac{1}{2} \dot{m}_p V_{exit}^2 = (\eta_F \eta_h \eta_N \dot{Q}_F)$ = fraction converted in propulsion power

Due to the minor mass of the Hydrogen propellant, at the same power regimen and at the same temperature an NTR/NTP will have higher specific impulse. In fact:

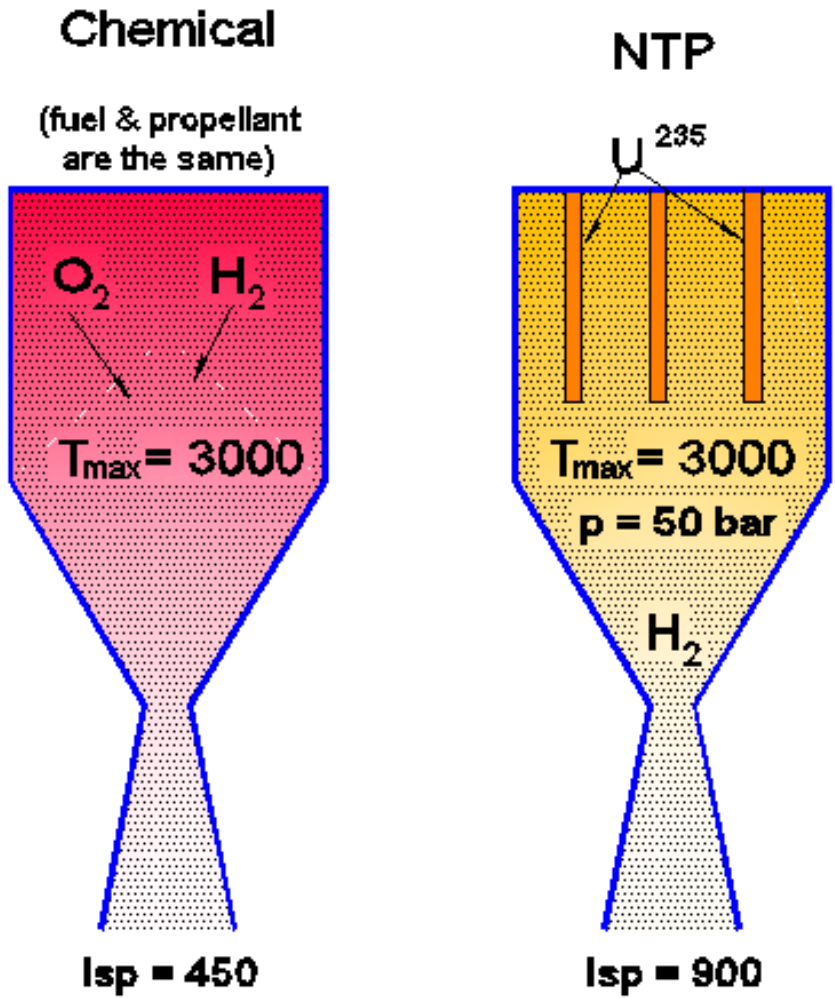

and furthermore at a very lower propellant pressure in the engine. A family of modular bimodal NTR (BNTR) space-transfervehicle concepts has been developed by NASA. It utilizes a common core stage powered by three ~70 kN~15-klbf engines that produce 50 kW of total electrical power for crew life support, high-data-rate communications with Earth, and an active refrigeration system for long-term, zero-boiloff liquid hydrogen ($LH_2$) storage. Candidate nuclear fuels for BNTR engines include uranium carbide ($UC_2$) particles with a chemical vapor-deposited coating in graphite and uranium-carbide zirconium-carbide (UC-ZrC) in graphite, which were developed during the Nuclear Engine for Rocket Vehicle Application (NERVA) program, as well as uranium oxide ($UO_2$) in tungsten (W) metal cermet. These fuels, which are listed in order of increasing temperature capability, can produce hot hydrogen exhaust ranging from ~2550 to 2800 K. Each engine has its own closed-cycle Brayton rotating unit, capable of generating up to 25 kW that provides an engine-out capability. Under nominal conditions, each Brayton rotating unit would operate at two-thirds of the rated power (~17 kW).

Thanks to the large amount of energy, during the flight artificial gravity can be generated into the BNTR vehicle to ensure crew health and fitness on long missions. The crew transfer vehicle could rotate (for example at ~4 rpm to provide the crew with a Mars gravity environment (~0.38 times Earth) during the outbound transit. A higher rotation rate of ~6 rpm would provide about 0.8g during the return leg of the mission to help reacclimatize the crew to Earth's gravity. A variant of the LH2 NTR option, known as the LOx-augmented NTR (or LANTR), would add an oxygen afterburner nozzle to the BNTR if variable thrust, variable Isp, or stage volume reduction was needed.

Even stronger perspectives are connected to the possibility of using exotic nuclear fuels and exotic reactor design as that, firstly proposed by Carlo Rubbia, of using very thin layers of 242 americium as nuclear fuel in order to obtain a direct energy transfer from the fission fragments (FF) to the hydrogen. Such a kind of reactor engine could have an engine temperature (~2000 K) lower than that of the propellant gas (>3000 K) and a very high specific impulse of the order of 1400 seconds (usually the specific impulse is

definited for rockets by the equation:

$$T_{hrust} = I_{sp} \frac{\Delta m}{\Delta t} g$$

where g is the gravitational accelerationso that the unit results to be seconds instead of Ns/Kg) In fact the direct transfer of energy from the FF to the gas that is quickly ejected from the nozzle allows to prevent the heating of the engine by the hot propellant gas that is at an extremely low pressure. Thus the direct conversion from kinetic energy of the FF to propellant enthalpy allows to operate a relatively cold engine (Te~2000K) with a very hot propellant ($T_P$>>3000 K). The specific impulse grows with the square root of the temperature.

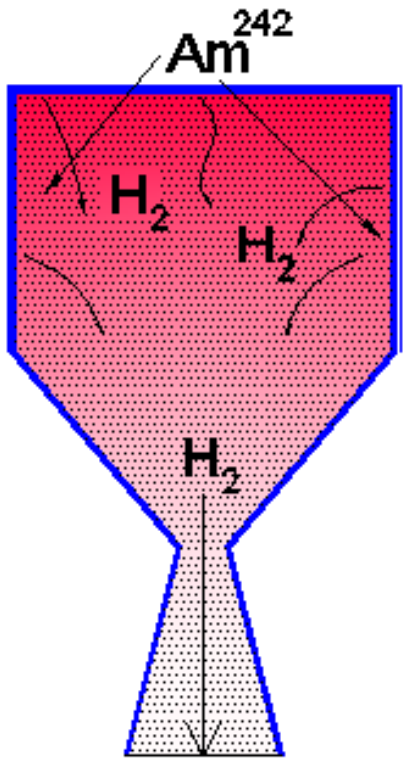

The FF engine would be much more efficient for both the use of propellant and temperature of the gas.
In the following figures it is shown the propellant consumption and the specific impulse for several types of engines.

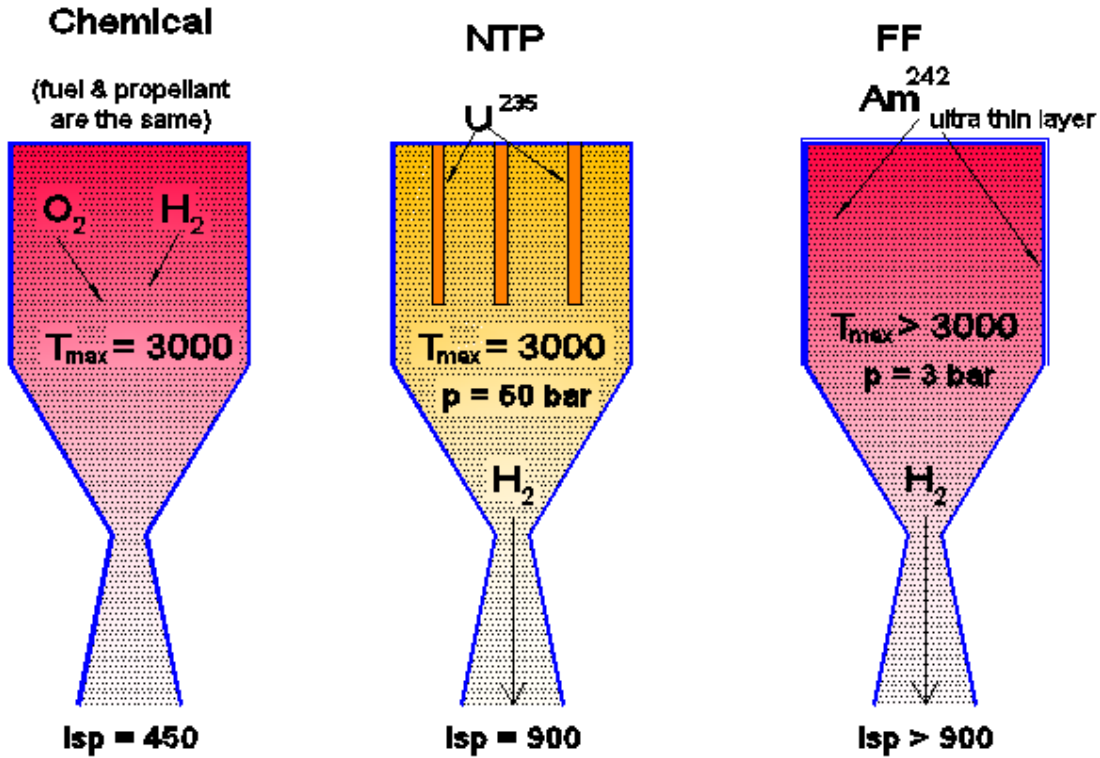

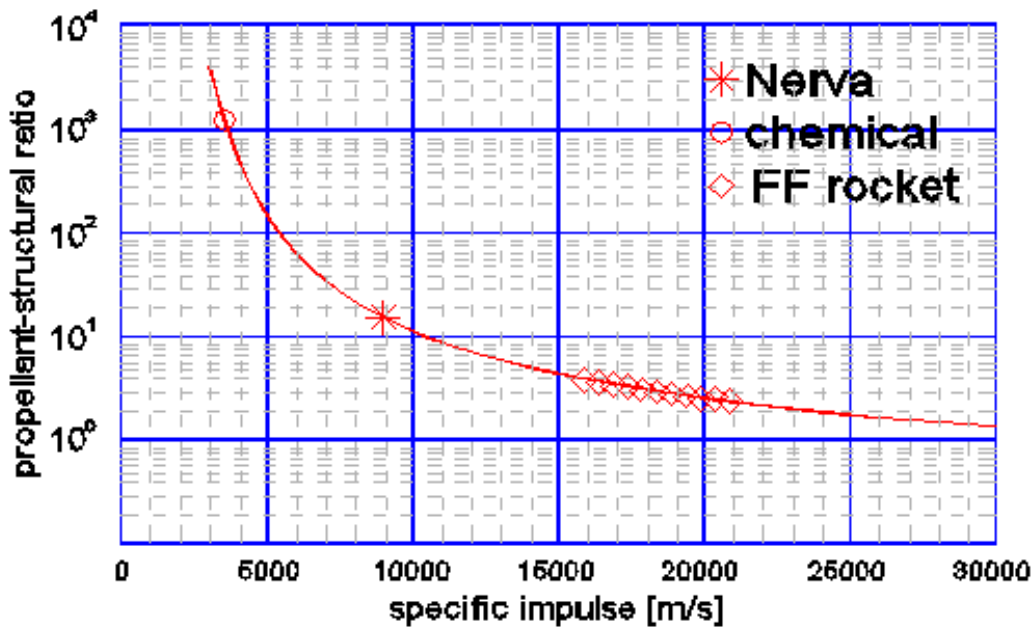

Hybrid reactor design is of course possible in order to optimize the performances to the specific mission; fuel preserving configurations for very long (~30/50 years) deep space missions, faster burning for manned missions to Mars or giants planets satellites.

Conclusions

The exploration of the interstellar medium and of the external heliosphere is a must in order to understand the origin and final fate of the solar system. Extremely interesting results can also be obtained by manned missions on Mars and the satellites of Jupiter and Saturn. Both these kind of missions are impossible without a revolution in the propulsion system of the spacecrafts. Omitting the quite rough option of an Orion like spacecraft, both nuclear reactor stoked up Ion thrusters and NTR propulsion seem interesting options; the first only for very light, very long deep space missions, the second, in various options, both for heavy multiprobe deep space investigations and for shorter planetary manned missions.